\documentclass[10pt]{article}
\usepackage{amsmath,amssymb,amsthm}
\usepackage[left=2cm,right=2cm,top=1.5cm,bottom=1.5cm]{geometry}
\usepackage{graphicx}

\title{Application of Bootstrap Re-sampling Method to a Categorical Data of HIV/AIDS Spread across different Social-Economic Classes}
\author{Bello A.O.$^{1}$, Oguntolu F. A.$^{2}$, Adetutu O. M.$^{3}$, J. P. Ojedokun$^{4}$\\
Department of Mathematics and Statistics$^{1,2,3}$\\
Federal University of Technology Minna,Nigeria.\\
\vspace{0.05in}
oyedele.bello@futminna.edu.ng$^{1}$\\
Department of Statistics$^{4}$\\
University of Ibadan, Nigeria.
}
\date{}

\theoremstyle{plain}

\begin{document}
\maketitle
\setlength{\parindent}{0pt}

\begin{abstract}
\noindent
 This research reports on the relationship and significance of social-economic factors (age, sex, employment status) and modes of HIV/AIDS transmission to the HIV/AIDS spread.  Logistic regression model, a form of probabilistic function for binary response was used to relate social-economic factors (age, sex, employment status) to HIV/AIDS spread. The statistical predictive model was used to project the likelihood response of HIV/AIDS spread with a larger population using 10,000 Bootstrap re-sampling observations. 
 \end{abstract}

\section{Preliminaries}
\ \ A virus has been draining the resources of the world for the past three decades and the virus is called HIV, which stands for Human Immunodeficiency Virus. The virus damages the immune system after a period of time, and this causes a variety of symptoms known as AIDS (Acquired immune deficiency syndrome). According to UNAIDS/HIV [13] “the pandemic of HIV and AIDS has continued to constitute serious health and socio economic challenges globally, for more than three decades. In the underdeveloped and developing countries which includes Nigeria. HIV/AIDS has reversed many of the health and developmental gains over the past three decades as reflected by indices such as life expectancy at birth and infant mortality rate among others”.\\
\par The interest of this work is to model the relationship pattern that describes the way at which HIV infection varies across gender, economic status and age-groups. Also, to determine the notable factor(s) that play major role in transmitting the virus among people of different social and economic classes per given population, and out of the various possible mode of transmission (PMOT), age-group, gender and economic status; we want to determine if there is (are) any existence of individual or collective variability effect(s) of these variables to HIV infection or otherwise. We shall verify the stability of our predictive model in order to obtain a prototype models that will adequately describe the pattern of the spread and also predict the trends of HIV infection among person(s) of different socio-economic classes in Oyo state. This will help the state government and other relevant non-governmental bodies to know the age group(s) or socio-economic class that required urgent or long term plan interventions.\\
 \footnote{ Manuscript* of the corresponding author: oyedele.bello@futminna.edu.ng (A. O. Bello) Published online at http://journal.sapub.org/statistics Copyright © 2015 Scientific \& Academic Publishing. All Rights Reserved}

\subsection{Epidemiology of HIV and AIDS} HIV infects cells in the immune system and the central nervous system. The main type of cell that HIV infects is the T helper lymphocyte; these cells play a crucial role in the immune system by coordinating the actions of other immune system cells. A large reduction in the number of T helper cells seriously weakens the immune system as HIV infects the T helper cell through the protein CD4 on its surface. HIV produces new copies of itself, which can then go on to infect many other cells. This process usually takes several years, see [2] for details. HIV virus is a communicable infection, for this work we classified all possible mode of transmission (PMOT) of HIV under the following medically confirmed activities: Sexual intercourse/ heterosexual contact, Pregnancy (mother-child transmission), Sharp objects and Blood transfusion. Also, the social status is classified as Employed, and Unemployed.
According to Open Dictionary[10] Employment is work that you are paid regularly to do for a person or company while Unemployed is define according to International Labour Organization[4] as active people who are without jobs.

\section{Literature Review}
The logistic regression model is one of the popular statistical models for the analysis of binary data with applications in physical, biomedical, behavioural sciences, and many others. Logistic regression analysis was implemented to determine the significant contributory factors influencing the subject of study [9]. The cases having the response variable as categorical, often called binary of (yes/no; present/absent; etc) and possible explanatory variables which can either be categorical variables, numerical variables or both are numerous in the biometry, psychometric, and epidemiology researches. In a longitudinal study of coronary heart disease as a function of age, gender, smoking history, cholesterol level, percentage of ideal body weight, and blood pressure, the response variable $y_i$ was defined to have the two possible outcomes: person developed heart disease during the study or person did not develop heart disease during the study were modelled using the logistic regression model (See [9] p555-556).\\
\ \\
 L.M Raposo and et al [7] used the logistic regression model to predict resistance to HIV protease Inhibitor, the model obtained was said to be useful in decision making regarding the best therapy for HIV positive individuals. Also, Jinma Ren and et al [6] {\it “Risk of Using logistic Regression to illustrate exposure-response relationship of infectious diseases”, the work was suspicious of the suitability of ordinary, categorical exposures, and logarithm transformation functions presented in logistic regression model to assess if the likelihood of infectious diseases is risk or as a result of exposure using simulated data.} {\bf However, the risk of using logistic regression is no risk at all if large sample size is used or procedure of large sample technique such as bootstrap re sampling method is used, this will reduce the bias in our estimates} {\it \bf as it shall be demonstrated in this work}. “The odd function is the most suitable function for interpretation of binary predictive problems” [1]. \footnote{ Manuscript* of the corresponding author: oyedele.bello@futminna.edu.ng (A. O. Bello) Published online at http://journal.sapub.org/statistics Copyright © 2015 Scientific \& Academic Publishing. All Rights Reserved}
 
\subsection{Data Description and Limitation} The data used in this study can be classified as secondary data because they were not generated by the investigator. Secondary data is a data collected initially for a particular purpose and it may not always provide detailed information, which a researcher needs. The implication of this kind of problem is that the researcher will generally resort to certain assumptions so as to fill the missing information. This reduces the scope, quality and amount of information required for the research. Information on different variables of interests was collected from the records of Central Blood Transfusion Service Unit Oyo State, Nigeria. Information collected is sub topic into the followings data variables:
Gender (Qualitative Variable): Male/ Female, Age (Quantitative Variable), Possible Mode of Transmission (PMOT), Employment Status (Emp): Employed / Unemployed. 400 observations of people that randomly visit the centre for HIV test for the purpose of medical diagnosis or for the purpose of blood donation were extracted out of the record that covers between years 2009 to 2014.\\
\ \\
 The dataset includes the Age of people that took the HIV test, so it is possible to calculate age in single years or age-groups. Presence of HIV (HIV+) and Absent of HIV (HIV -) is measured as a simple dichotomy coded one and zero respectively. The fact that we treat all predictors as discrete factors allows us to summarize the data in terms of the numbers of HIV+ and HIV- in each of the five different age-groups. The reference classes of explanatory variables (the male and the employed population) are coded 0, because they are traditionally believed to be less susceptible to HIV infection. We cross tabulate the variable PMOT (sexual intercourse, sharp object, mother-to-child and Blood transfusion) against age-group (0-15, 16-39, 40-54, 55-69, 70 and above in years), this will be treated as a qusi-experimental data.
\section{Methodology}

\noindent 
Considering the case where our response $y_{i} $ is a dichotomous response, when possible response is either yes or no, death or alive, present or absent and as the case at hand in this work is either HIV negative or HIV positive.
$$y_{i} =\left\{\right. _{0}^{1} {}_{if{\rm \; ith\; individual\; is\; HIV\; negative}}^{if{\rm \; ith\; individual\; is\; HIV\; positive}} $$
We code the present or absent of subject of study as 1 or 0 respectively. The distribution of $y_{i} $ is binomial of a single trial or basically Bernoulli distribution as used by some text. The binary indicator variable outcome can only be 1 or 0 as the probability is bound between 0 and 1; this gives a sigmodial shape approaching 0 and 1 asymptotically. This is a nonlinear problem. The logistic regression is suitable for such problem usually when response variable is qualitative of two possible outcomes.

\noindent The logistic function relating $y_{i} $ to predictors which can be qualitative, quantitative or both is a very flexible model which makes it vital to solving many epidemiology and social indicator related problems. The logistic line can also be of form;

\begin{equation}
y_{i} =\theta _{0} +\theta _{1} x_{1} +\theta _{2} x_{2} +...+\theta _{k} x_{k} \ \ \ \ \  Where\ \    y_{i} =0, 1  
\end{equation}
\begin{equation} 
E(y_{i} )=\theta _{0} +\theta _{1} x_{1} +\theta _{2} x_{2} +...+\theta _{k} x_{k}
\end{equation} 
Probabilities; given $p(y_{i} =1)=H_{i} $ and $p(y_{i} =0)=1-H_{i} $, therefore
\begin{equation} \label{GrindEQ__3_} 
E(y_{i} )=1(H_{i} )+0(1-H_{i} )=H
\end{equation} 
$H$ is probability of our subject of interest in study taking place and $1-H$ is the probability of subject of interest not occurring. The subject of interest informs our choice of coding 1 or 0, in the case of an indicator variable such as we have HIV+ is coded 1 and HIV- coded 0.

\noindent Equation \eqref{GrindEQ__3_} gives the probability $y_{i} =1$ given that level of parameter variable is $X_{i}$. Logistic regression model is a special case of general linear model, only that its conditional probability follows a Bernoulli distribution. The special problems associated with model having binary response variable is the problem of having our error terms not normally distributed and heteroskadastic in nature due to the distribution of our response variable bonded between 0 and 1. 

          $$\left. \begin{array}{l} {e_{i} =y_{i} -(\theta _{0} +\theta _{1} x_{1} +\theta _{2} x_{2} +...+\theta _{k} x_{k} ){\rm \; \; }} \\ {y_{i} =0;{\rm \; \; \; }\varepsilon =1-\theta _{0} -\theta _{1} x_{1} -\theta _{2} x_{2} -...-\theta _{k} x_{k} } \\ {y_{i} =0;{\rm \; \; \; }\varepsilon =-\theta _{0} -\theta _{1} x_{1} -\theta _{2} x_{2} -...-\theta _{k} x_{k} } \end{array}\right\}\eqno(3.1)$$

\noindent $\varepsilon $  is not normally distributed. Other problem associated with the logistic model is the constraints condition on response function (See Michael H. Kutner and et al; \textit{Applied Linear Statistical Models} Fifth Edition, p557-558).

\noindent The function form in equation \eqref{GrindEQ__3_} has its left hand-side take value ranging between 0 and 1, while the right hand-side is not in a form that can return values between 0 and 1 asymptotically. Therefore, we require a link function to properly link the left hand-side to the right hand-side. Link function such as identity will not be appropriate for the initial nonlinear problem at hand. However, for easier understanding and interpretations the logit function is usually employed. The model is initially best put in the form;

\begin{equation}
(H)=\log \frac{H}{1-H} 
\end{equation}
\noindent Also;
\begin{equation} \label{GrindEQ__5_} 
\log \frac{\hat{H}(x)}{1-\hat{H}(x)} =\theta _{0} +\theta _{1} x_{1} +\theta _{2} x_{2} +...+\theta _{k} x_{k}   
\end{equation} 
The interpretation of $\theta 's$  is not straightforward because increase in unit of X varies for the logistic regression model according to the location of the starting point of the X scale (Michael H. Kutner and et al). The logit function is the natural logarithm (In) of odds of $y$ and taking exponential of the log of odd function gives us the most appreciable odd function, vital in our interpretation of result. The odd function will simplify our interpretation problem.
\begin{equation} \label{GrindEQ__6_} 
\frac{H}{1-H} =e^{\theta _{0} +\theta _{1} x_{1} +\theta _{2} x_{2} +...+\theta _{k} x_{k} } 
\end{equation} 
Explicitly;
\[H=(1-H){\rm \; }e^{\theta _{0} +\theta _{1} x_{1} +\theta _{2} x_{2} +...+\theta _{k} x_{k} } \] 
\[H=(e^{\theta _{0} +\theta _{1} x_{1} +\theta _{2} x_{2} +...+\theta _{k} x_{k} } -H{\rm \; }e^{\theta _{0} +\theta _{1} x_{1} +\theta _{2} x_{2} +...+\theta _{k} x_{k} } ){\rm \; }\] 
\[H+H{\rm \; }e^{\theta _{0} +\theta _{1} x_{1} +\theta _{2} x_{2} +...+\theta _{k} x_{k} } =e^{\theta _{0} +\theta _{1} x_{1} +\theta _{2} x_{2} +...+\theta _{k} x_{k} } {\rm \; }\] 
\[H(1+{\rm \; }e^{\theta _{0} +\theta _{1} x_{1} +\theta _{2} x_{2} +...+\theta _{k} x_{k} } )=e^{\theta _{0} +\theta _{1} x_{1} +\theta _{2} x_{2} +...+\theta _{k} x_{k} } {\rm \; }\] 
\[H=e^{\theta _{0} +\theta _{1} x_{1} +\theta _{2} x_{2} +...+\theta _{k} x_{k} } \frac{1}{(1+{\rm \; }e^{\theta _{0} +\theta _{1} x_{1} +\theta _{2} x_{2} +...+\theta _{k} x_{k} } )} {\rm \; }\] 
\begin{equation} \label{GrindEQ__7_} 
H=\frac{e^{\theta _{0} +\theta _{1} x_{1} +\theta _{2} x_{2} +...+\theta _{k} x_{k} } }{(1+{\rm \; }e^{\theta _{0} +\theta _{1} x_{1} +\theta _{2} x_{2} +...+\theta _{k} x_{k} } )} {\rm \; }
\end{equation} 
The inverse of the logit function is the logistic function.

\noindent Hence; 
\begin{equation} \label{GrindEQ__8_} 
H_{i}=probability{\rm \; }(0,1/X=x)=\frac{e^{\theta _{0} +\theta _{1} x_{1} +\theta _{2} x_{2} +...+\theta _{k} x_{k} } }{(1+{\rm \; }e^{\theta _{0} +\theta _{1} x_{1} +\theta _{2} x_{2} +...+\theta _{k} x_{k} } )} {\rm \; }
\end{equation} 
The logistic function form will return the right hand-side to be property value ranging from 0 and 1. The function increases monotonically if the gradient $\theta >0$ and decreases monotonically if $\theta <0$.

\noindent Algebraically the equation 7 or 8 is also of the form in equation \eqref{GrindEQ__9_};
\[\frac{\exp (\theta _{0} +\theta _{1} x_{1} +\theta _{2} x_{2} +...+\theta _{k} x_{k} }{[1+\exp (-\theta _{0} -\theta _{1} x_{1} -\theta _{2} x_{2} -...-\theta _{k} x_{k} )]} \] 
\[[\frac{1}{\exp (\theta _{0} +\theta _{1} x_{1} +\theta _{2} x_{2} +...+\theta _{k} x_{k} } ]^{-1} [1+\exp (\theta _{0} +\theta _{1} x_{1} +\theta _{2} x_{2} +...+\theta _{k} x_{k} ]^{-1} \] 
\[[\frac{1}{\exp (\theta _{0} +\theta _{1} x_{1} +\theta _{2} x_{2} +...+\theta _{k} x_{k} } +1]^{-1} \] 
\begin{equation} \label{GrindEQ__9_} 
E(y_{i} )=H_{i} =[1+\exp (-\theta _{0} -\theta _{1} x_{1} -\theta _{2} x_{2} -...-\theta _{k} x_{k} )]^{-1}  
\end{equation} 
\subsection{Method of Estimation}

\noindent The variability of the error terms variances differs at different level of X, as shown in equation (3.1) . This makes the ordinary least square estimation ineffective in estimation of logistic function. The maximum likelihood is a better method for estimating logistic function since logistic function predicts probabilities, and not just classes, it can fit the probabilities for each class of our data-point, either for the class $'H_{i{\rm \; }} 'or{\rm \; \; '1-H}_{{\rm i}} {\rm '}$. We must also note that the error term is not usually considered in logistic problems.\textbf{}

\noindent 

\subsubsection{Maximum Likelihood Estimation }

\noindent 

\noindent  The maximum likelihood estimate is that value of the parameter that makes the observed data most likely [12]. The values of $\theta $s  that maximize $\log _{e} L(\theta )$, that is, the value of $\theta $ that assign the highest possible probability to the sample that was actually obtained.  The method of likelihood in estimating a logistic function usually requires numerical procedures, and Fisher scoring or Newton-Raphson which often work best. Most statistical packages have the logit numerical search procedure. In this work,  R-programming language package for logistic regression for obtaining the maximum likelihood estimates of a logistic regression is used.
 \footnote{ Manuscript* of the corresponding author: oyedele.bello@futminna.edu.ng (A. O. Bello) Published online at http://journal.sapub.org/statistics Copyright © 2015 Scientific \& Academic Publishing. All Rights Reserved}

\noindent     Let $y_{1,} y_{2,...} y_{k,} $ be n independent random variables (r.v.'s) with probability density functions $f(y;\theta )$ that depends on parameter $\theta $ . The likelihood of the joint density function of k independent observations is $L(\theta )=f(y_{1} ,y_{2} ,...y_{k;} \theta )$. Then;
\begin{equation} \label{GrindEQ__10_} 
f(y;\theta )=\mathop{\prod }\limits_{i=1}^{n} f_{i} (y_{i} ;\theta )=L(\theta ;y)
\end{equation} 
The root of the equation is obtained by equating the first derivative of equation \eqref{GrindEQ__10_} to zero and the maximum likelihood estimate (MLE) hold when the second derivative is negative.

\noindent 

\noindent The probability distribution function of our $y_{i} $ follows the Bernoulli distribution,$y_{i} =H_{(x_{i} )} ^{y_{i} } (1-H_{(x_{i} )} )^{1-y_{i} }$ with $y_{i} $ taking zero or one. The likelihood function is;

\noindent   
\[L(\theta _{i} )=\mathop{\prod }\limits_{i=1}^{n} H_{i_{} } ^{y_{i} } (1-H_{i_{i} } )^{1-y_{i} } \] 
\[\mathop{\prod }\limits_{i=1}^{n} \left(H_{(x_{i} )} \right)^{y_{i} } (1-H_{(x_{i} )} )^{-y_{i} } _{} (1-H_{(x_{i} )} )^{1} \]

\begin{equation} \label{GrindEQ__11_} 
\mathop{\prod }\limits_{i=1}^{n} \left(\frac{H_{(x_{i} )} }{1-H_{(x_{i} )} } \right)^{y_{i} } (1-H_{(x_{i} )} )^{} 
\end{equation}

Recall eqn \eqref{GrindEQ__7_} H is substituted into equation \eqref{GrindEQ__11_}
\[{\rm \; } \mathop{\prod }\limits_{i=1}^{n} \left(\frac{\frac{e^{\theta _{0} +\theta _{1} x_{1} +\theta _{2} x_{2} +...+\theta _{k} x_{k} } }{(1+{\rm \; }e^{\theta _{0} +\theta _{1} x_{1} +\theta _{2} x_{2} +...+\theta _{k} x_{k} } )} }{[1-\frac{e^{\theta _{0} +\theta _{1} x_{1} +\theta _{2} x_{2} +...+\theta _{k} x_{k} } }{(1+{\rm \; }e^{\theta _{0} +\theta _{1} x_{1} +\theta _{2} x_{2} +...+\theta _{k} x_{k} } )} ]_{} } \right)^{y_{i} } [1-\frac{e^{\theta _{0} +\theta _{1} x_{1} +\theta _{2} x_{2} +...+\theta _{k} x_{k} } }{(1+{\rm \; }e^{\theta _{0} +\theta _{1} x_{1} +\theta _{2} x_{2} +...+\theta _{k} x_{k} } )} ]^{} \]

\[\mathop{\prod }\limits_{i=1}^{n} \left(\frac{\frac{e^{\theta _{0} +\theta _{1} x_{1} +\theta _{2} x_{2} +...+\theta _{k} x_{k} } }{(1+{\rm \; }e^{\theta _{0} +\theta _{1} x_{1} +\theta _{2} x_{2} +...+\theta _{k} x_{k} } )} }{\frac{1+e^{\theta _{0} +\theta _{1} x_{1} +\theta _{2} x_{2} +...+\theta _{k} x_{k} } -e^{\theta _{0} +\theta _{1} x_{1} +\theta _{2} x_{2} +...+\theta _{k} x_{k} } }{1+{\rm \; }e^{\theta _{0} +\theta _{1} x_{1} +\theta _{2} x_{2} +...+\theta _{k} x_{k} } } _{} } \right)^{y_{i} } (\frac{1+e^{\theta _{0} +\theta _{1} x_{1} +\theta _{2} x_{2} +...+\theta _{k} x_{k} } -e^{\theta _{0} +\theta _{1} x_{1} +\theta _{2} x_{2} +...+\theta _{k} x_{k} } }{1+{\rm \; }e^{\theta _{0} +\theta _{1} x_{1} +\theta _{2} x_{2} +...+\theta _{k} x_{k} } } )^{} \] 
\begin{equation} \label{GrindEQ__12_} 
\mathop{\prod }\limits_{i=1}^{n} \frac{(e^{\theta _{0} +\theta _{1} x_{1} +\theta _{2} x_{2} +...+\theta _{k} x_{k} } )^{y_{i} } }{1+{\rm \; }e^{\theta _{0} +\theta _{1} x_{1} +\theta _{2} x_{2} +...+\theta _{k} x_{k} } } ^{} 
\end{equation} 

Taking the natural logarithm
\begin{equation} \label{GrindEQ__13_} 
l(\theta )=\sum _{i=1}^{n}y_{i} {\rm \; }\theta _{0} +\theta _{1} x_{1} +\theta _{2} x_{2} +...+\theta _{k} x_{k}  -\sum _{i=1}^{n}{\rm In(1}+e^{\theta _{0} +\theta _{1} x_{1} +\theta _{2} x_{2} +...+\theta _{k} x_{k} } ) 
\end{equation} 
\[\begin{array}{l} {l(\theta )=\sum _{i=1}^{n}y_{i} {\rm \; (}\theta _{0} +\theta _{1} x_{1} +\theta _{2} x_{2} +...+\theta _{k} x_{k} ) -n_{i} .In(1+e^{\theta _{0} +\theta _{1} x_{1} +\theta _{2} x_{2} +...+\theta _{k} x_{k} } )} \\ {l(\theta )=\sum _{i=1}^{n}y_{i} {\rm \; (}\sum _{{\rm j}={\rm 0}}^{{\rm k}}{\rm x}_{{\rm ij}}  \theta _{k} ) -n_{i} .In(1+e^{\sum _{k=0}^{k}x_{ik}  \theta _{k} } )} \end{array}\] 

\[\frac{\partial l(\theta )}{\partial \theta _{j} } =\sum _{i=1}^{n}y_{i} x_{ij}  -\sum _{i=1}^{n}\frac{e^{\theta _{0} +\theta _{1} x_{1} +\theta _{2} x_{2} +...+\theta _{k} x_{k} } }{1+e^{\theta _{0} +\theta _{1} x_{1} +\theta _{2} x_{2} +...+\theta _{k} x_{k} } }  x_{ij} \] 
Recall equation \eqref{GrindEQ__7_} and substitute for$H$; the probability of subject of interest under study occurring

\begin{equation}
\frac{\partial l(\theta )}{\partial \theta _{j} } =\sum _{i=1}^{n}y_{i} x_{ij}  -\sum _{i=1}^{n}H_{i}  x_{ij} for\  \ j=1,2,...
\end{equation}

The differentiation of the log likelihood function in equation (13) with respect to each parameter  $\theta _{j} $ will not analytical give us the maximum likelihood estimates by setting each of the k equations in equation( 13) equal to zero. It is a system of k nonlinear equations. The solution to the K unknown variables is a nonlinear problem cannot be solved analytically but through numerical estimation using an iterative process. The Newton-Raphson method is popularly used for a logistic nonlinear function. However, problem of multicollinarity may arise which is visible when there are large estimated parameters and large standard error values. Also, convergence problem in numerical search procedure can be associated with multicollinearity problem which can be overcome by reducing the number of parameter variables for easy and quick convergence. For details see [8] and [9].

 \footnote{ Manuscript* of the corresponding author: oyedele.bello@futminna.edu.ng (A. O. Bello) Published online at http://journal.sapub.org/statistics Copyright © 2015 Scientific \& Academic Publishing. All Rights Reserved}

\subsection{Variance Estimation Of A Logistic Function Using The Bootstrap Method}
\noindent The general linear model rely on asymptotic approximations in estimating the coefficient standard errors and this may not be reliable, just as measures such as R-square based, residual errors are not very informative and can be misleading. Therefore, using the method of bootstrap (a re-sampling technique) will either confirm or dispel our doubts about the sufficiency of our sample to estimate unbiased and robust estimates for the population parameters. For our models to adequately capture the reality of HIV/AIDS spread across different socio-economical classes in Oyo state population as likely as possible, we shall generate 10,000 Bootstrap samples from the original sample to estimate our models’ parameter values and their confidence intervals. In addition to the bootstrap method we shall also consider the multiple split sample procedure. These will help us in selecting robust parameter values for our models. Bootstrapping technique has being identified to be effective in dealing with non-linear data with extremely non-normal distribution. See [3].
\subsection{The Odd Function}

J. M Bland and Douglas G [5] mentioned that there are mainly three reasons to use the odds ratio. “Firstly, they provide an estimate (with confidence interval) for the relationship between two binary variables. Secondly, they enable us to
162 A. O. Bello et al.: Application of Bootstrap Re-Sampling Method to a Categorical
Data of HIV/AIDS Spread across Different Social-Economic Classes
examine the effects of other variables on that relationship, using logistic regression. Thirdly, they have a special and very convenient interpretation.” The odds are nonnegative, with odds 1.0 when a success is more likely than a failure. According to Pedhazur [11] Odds are determined from probabilities and range between 0 and infinity. Odds are defined as the ratio of the probability of success and the probability of failure.\\
\ \\
 The odds of success given as $\frac{H}{1-H}$ and the odds of failure would be odds (failure) given as $\frac{1-H}{H}$. The odds of success and the odds of failure are just reciprocals of one another. Probability and odds both measure how likely it is that our subject of interest will occur. Notably, the sign of the log-odds ratio indicates the direction of its relationship, the distinction regarding a positive or negative relationship in that of the odds ratios is given by which side of 1 the odd values fall on. Odd value 1 indicates no relationship, less than one indicates a negative relationship and greater than one indicates a positive relationship. However, in order to get an intuitive sense of how much things are changing, we need to get the exponential of the log-odds ratio, which gives us the odds ratio itself [1]. The odd ratio of the odd for x=1 to the odd of x=0 is \\

\noindent 

\noindent The odd ratio of the odd for x=1 to the odd of x=0 is

\noindent 
\[{\rm \; the\; odd\; ratio}=\frac{{\raise0.7ex\hbox{$ {\rm H(1)} $}\!\mathord{\left/{\vphantom{{\rm H(1)} {\rm 1-H(1)}}}\right.\kern-\nulldelimiterspace}\!\lower0.7ex\hbox{$ {\rm 1-H(1)} $}} }{{\raise0.7ex\hbox{$ H(0) $}\!\mathord{\left/{\vphantom{H(0) 1-H(0)}}\right.\kern-\nulldelimiterspace}\!\lower0.7ex\hbox{$ 1-H(0) $}} } \]

\noindent 
\[    =  \left(\frac{\frac{e^{\theta _{0} +\theta _{1}  +\theta _{2} +...+\theta _{k} } }{(1+ e^{\theta _{0} +\theta _{1} +\theta _{2} +...+\theta _{k} } )} }{\frac{1}{(1+e^{\theta _{0} +\theta _{1} +\theta _{2} +...+\theta _{k} } )}} \right)\div \left(\frac{\frac{e^{\theta _{0} } }{(1+e^{\theta _{0} } ) }}{\frac{1}{(1+e^{\theta _{0} } )}  } \right) \]

\[                   =          \frac{e^{\theta _{0} +\theta _{1}  +\theta _{2} +...+\theta _{k} } }{e^{\theta _{0} } } \] 
\begin{equation}
         The\ \  odd \ \ ratio    ={\rm exp(}\theta _{{\rm 1}} )*{\rm \; }\exp (\theta _{{\rm 2}} ){\rm \; }*{\rm ....\; }*\exp (\theta _{{\rm k}} )
\end{equation} 

This result obtained is the relationship between the odds ratio and an independent dichotomous. The result tells that the odds on the event that y equals 1, increases (or decreases) by the factor $\exp (\theta _{1} +\theta _{2} +...+\theta _{k} )$ among those with x= 1 than among those x= 0. One major condition to note when interpreting for multiple logistic regression is that the estimated odds ratio for predictor variable x assumes that all other predictor variables are held constant.

\section{Results and Discussions}
\includegraphics*[width=6in, height=2.9in, keepaspectratio=false]{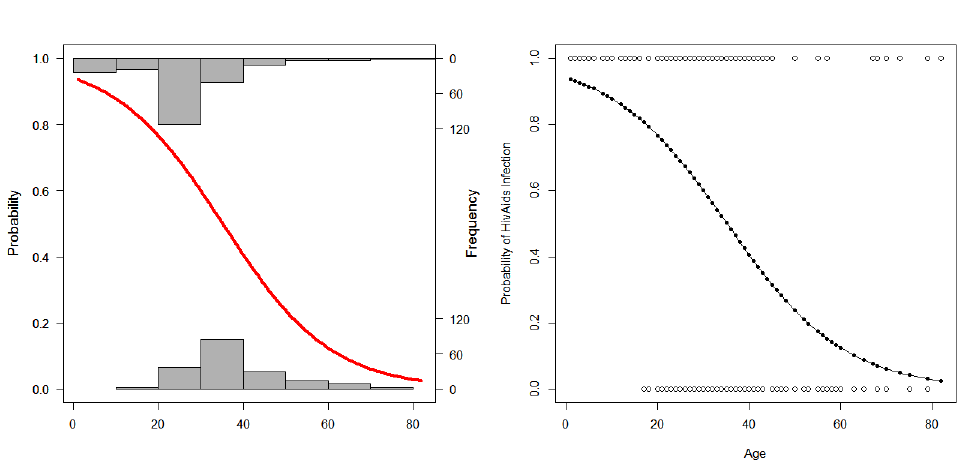}\begin{center}
 {Figure 1 \& 2: Exploratory Data Analysis for our data}
  \end{center}
  \newpage
\noindent A lowess nonparametric response curve was fitted for the data (see Figure 1 and 2), the plots show a sigmoidal S-shaped response function, and the lowess fit S-shape supported our choice of fitting a logistic regression model to the data.

\subsection{The Large-Sample Inference Procedure Results}

\noindent \textbf{}

\noindent I.\textbf{ Bootstrap Method Results and the Original Observation Results}
\begin{center}
\begin{tabular}{|p{0.7in}|p{1.0in}|p{0.8in}|p{0.8in}|p{1.0in}|p{0.7in}|p{0.8in}|} \hline 
\textbf{Parameters} & \textbf{\begin{center}Original observation estimates\end{center}} & \textbf{Log Odd C.I.\newline  2.5 \%   \newline 97.5 \%} & \textbf{Bootstrap Estimate} & \textbf{Bootstrap  C.I.\newline  2.5 \%   \newline 97.5 \%} & \textbf{   Odd\newline  Estimate} & \textbf{Odd C.I.\newline 2.5 \%   \newline 97.5 \%} \\ \hline 
\textbf{}\textbf{Intercept} & 1.56097 & 0.68127228  2.49950097 & 1.56096913 & 0.447\newline 2.623 & 4.7634354 & 1.9763907\newline 12.176416 \\ \hline 
\textbf{Age} & -0.07492 & -0.09838407 -0.05382945 & -0.074924 & -0.105\newline  -0.047 & 0.9278142    & 0.9063008  0.9475937 \\ \hline 
\textbf{Emp} & 1.64392 & 1.16798036  2.13580532 & 1.64391607 &  1.141\newline  2.147 & 5.1753971    & 3.2154919  8.4638598 \\ \hline 
\textbf{Gender} & 0.08356 & -0.42697826      0.58909633 & 0.08355517 & -0.4610  0.6125 & 1.0871452 & 0.6524777  1.8023589 \\ \hline 
\end{tabular}                                                  Table 1
\end{center}
\noindent \includegraphics*[width=6in, height=4in, keepaspectratio=false]{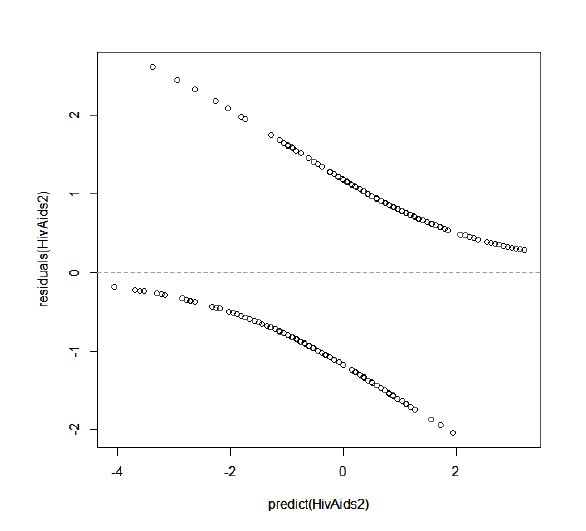}
                            Figure 3

\noindent{\bf I. Bootstrap Re sampling Technique:}
\par The numerical search converges after four iterations with a very small standard error for each parameter shown in table 1. The residuals plotted against the predicted probability (See figure 3), shows the lowess smooth approximates a line having zero slope and intercept, and we can conclude that model inadequacy is not apparent.\\
\par The original sample confidence intervals constructed for the coefficient estimates and that of the bootstrap confidence intervals coincide at almost the same intervals; they agree quite well and these demonstrate the precision of the model coefficient estimates. The parameter estimates from the original observation and the 10,000 bootstrap samples were asymptotically the same, thus, we can conclude within approximately 95 percent confidence that our sample size is as sufficient as using any other larger sample size, all of our coefficient estimate are between 2.5\% and 97.5\% respectively. (See table 1). Statistically, our sample size is a good representation of the entire population and sufficient to inferring the population characteristics.\\

 Also, from table 1 above, the coefficient of factors Age and Emp shows great statistical significance at 0.05 level of significance with a very small standard error values of 0.01133 and 0.246 respectively. Gender is significant but not at 95\% level of significance; however, because of the prior importance of this natural factor to HIV infections in area of sexual intercourse, and also its significance at interaction level with Employment status (Emp.), we shall retain the Gender coefficient in our model.\\
 
  The odds ratio of HIV infection for the employed as our reference class in Emp. variable and that of the male as reference class in variable Gender is exp (1.56097). The male gender has almost five times lesser odd of contracting HIV compared to the female. Moreover, value 1.56097 is the log odds ratio of male with employment of a given age contracting HIV; the employed male population in Oyo State odd ratio can also be translated to their probabilities of contracting HIV, given by exp(1.56097)(1+exp(1.56097) ) 0.8264925.\\
  
The negative coefficient value of Age parameter suggests a negative relationship between age and HIV infection, which imply that the probability of contracting HIV decreases as Age of person(s) increases. The odds of contracting HIV given age cannot be given a direct interpretation based on the question ‘what unit of age is appropriate and applicable to show the change in odds ratio?’ The odds is best described by $exp(c*Age)$, given c is a difference of units of ages under comparison. For the difference of unit between age 39 and 54years, the odds of contracting HIV between age 39 and 54years is $exp(15*(-0.07492))=0.3250423$. For this we can now say the odd of contracting HIV decreases by 33\% with each additional 15 years increase in age. The inverse relationship between age and Probability of HIV infection suggests that the younger generation below 55years should be of first priority in all the efforts towards eradicating HIV/AIDS spread.\\

 Also we could say that the result indicates a possibility that 33\% of the infected persons in Oyo state never survived the AIDS beyond 15years. This might be as a result of inadequate medical supports, psychological stigma and discrimination that is still associated with HIV/AIDS infection in Africa, Oyo state inclusive. This call for the attention of all relevant organisations to increase their support for HIV/AIDS victims, intensify campaigns against stigmatization, discrimination of HIV/AIDS patient and also give supports such as free or subsidize antiretroviral drugs.\\ Also, from the table 1 above, the odds of a unemployed person(s) in Oyo state contracting HIV are five times more likely than that of their employed counterpart and the generally probability of unemployed population of a given age contracting HIV is exp(1.56097+1.64392)/ (1+exp(1.56097+1.64392))=0.9610179. This result shows that the unemployed are more susceptible to HIV. \\
 
 The difference in the odds ratio of HIV infection between the female and the male individuals in the population is 1.087145(8.7\%). This result implies that the odds the females in the population contracting HIV are 8. 7\% than that of the males in the population, for given age and employment status. The positive coefficient of the gender and employment status (Emp.) variables imply that the female population in Oyo state are more likely to be HIV infected than the male counterpart and the unemployed population more likely that the employed respectively.
Recall

\noindent Recall$H_{i} =probability{\rm \; }(0,1/X=x)=[1+\exp (-\theta _{0} -\theta _{1} x_{1} -\theta _{2} x_{2} -...-\theta _{k} x_{k} )]^{-1} $;
\begin{equation} \label{GrindEQ___A2016_A20_} 
\hat{H}=[1+\exp (-1.56097+0.07492Age_{} -1.64392Emp_{} -0.08356Gender_{} )]^{-1} \dots \dots  
\end{equation}

\begin{enumerate}
\item [{II}]  \textbf{Multiple Split Sample Procedure Result}
\end{enumerate}
 \footnote{ Manuscript* of the corresponding author: oyedele.bello@futminna.edu.ng (A. O. Bello) Published online at http://journal.sapub.org/statistics Copyright © 2015 Scientific \& Academic Publishing. All Rights Reserved}

\noindent       

\noindent            glm(formula = HIV \~{} Age + Emp + Gender, family = binomial(link = "logit"), data = HivAids.dat2)                                                                                            

\noindent          50,100,200,300 and 400 
\[\left. \begin{array}{l} {Model{\rm \; for\; 50\; samples}} \\ {H_{i} =[1+\exp (-{\rm 1.35357}+{\rm 0.08088}Age_{} -{\rm 2.82061}Emp_{} +{\rm 0.23497}Gender_{} )]^{-1} } \\ {Model{\rm \; f}or{\rm \; 100\; samples}} \\ {H_{i} =[1+\exp (-{\rm 0.92883}+{\rm 0.06257}Age_{} -{\rm 2.28798}Emp_{} +{\rm 0.31093}Gender_{} )]^{-1} } \\ {{\rm Model\; for\; 200\; samples}} \\ {H_{i} =1+\exp (-{\rm 1.61271}+{\rm 0.07333}Age_{} -{\rm 2.16267}Emp_{} +{\rm 0.40551}Gender]^{-1} } \\ {Model{\rm \; for\; 300\; samples}} \\ {H_{i} =1+\exp (-{\rm 1.30131}+{\rm 0.06921}Age_{} -{\rm 1.67899}Emp_{} +{\rm 0.15188}Gender]^{-1} } \end{array}\right\}(17)\]

\noindent The equation (16) is from the 400 samples. Equation (17) was fitted from different samples of 50, 100, 200 and 300 data points respectively, their probability plots across different age is shown by figure 4; As the sample sizes increase the closer the probabilities tends asymptotically to a sure prediction. This confirmed the central limit theorem showing that as our sample size increases the closer we get to the true value of the unknown population parameter’s estimates and the closer to the true probability. Therefore using the parameter estimates obtained from the larger 10,000 bootstrap samples is not a bad idea.
\noindent 

\noindent \includegraphics*[width=6in, height=4in, keepaspectratio=false]{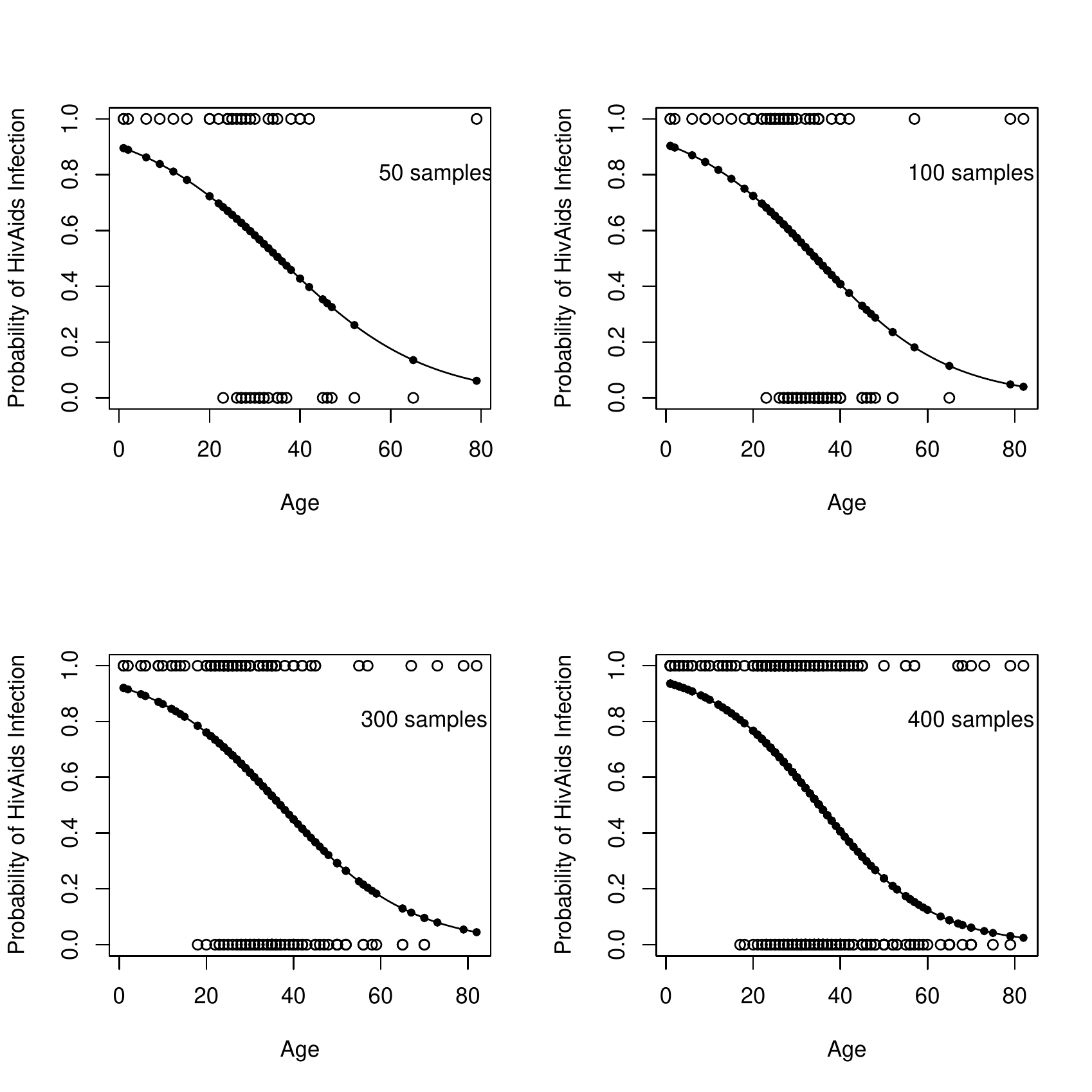}
                            Figure 4

\textbf{III. Model  Validation  }\\
\par \ \ \ \ \ \ \ \ \ \ \ Using model\rm \; for\; 300\; samples
\begin{equation}
  H_{i} =1+\exp (-{\rm 1.30131}+{\rm 0.06921}Age_{} -{\rm 1.67899}Emp_{} +{\rm 0.15188}Gender]^{-1} 
\end{equation}
\noindent The predictive model for Male-Employed individuals;
\begin{equation}
prob.logOdds.Male.Emp=exp(1.56097 -0.07492*Age)/ (1+ 1.56097 -0.07492*Age))
\end{equation}
The model prediction from the equation (18) above was obtained from the first 300 observations. We now used equation (18) to predict $H_i$ with $i=301 to 400$, within our observations, using predictor variables within our observation from data point 301 to 400. We compared the predicted result with the original observation of HIV from data point 301 to 400. The model gave correct prediction of 92 out of the 100 data points predicted. These imply that the fitted model is more than 91\% stable with less than 8 percents variations compared with the original data. Our model has not done badly. We can now at this confidence predict the likelihood of HIV spread in Oyo State among different gender, economic classes across all possible age.\\

\noindent The predictive model for Male-Unemployed individuals;
\begin{equation} 
prob.logOdds.Male.UnEmp   = \frac{ exp(1.56097\; -0.07492*Age+1.64392*1)}
{(1+exp\; (1.56097\; -\; 0.07492*Age\; \; \; +\; 1.64392*1))}
\end{equation} 
\noindent The predictive model for Female-Employed individuals;
\begin{equation}
 prob.logOdds.Female.Emp = \frac{\; exp\; (1.56097\; -0.07492*Age\; +\; 0.08356*1)}{ (1+\; exp(1.56097\; -0.07492*Age\; +\; 0.08356*1))} 
\end{equation}
\noindent 
 \footnote{ Manuscript* of the corresponding author: oyedele.bello@futminna.edu.ng (A. O. Bello) Published online at http://journal.sapub.org/statistics Copyright © 2015 Scientific \& Academic Publishing. All Rights Reserved}

\noindent The predictive model for Male-Unemployed individuals

\begin{equation}
 prob.logOdds.Female.UnEmp   =\frac{ exp(1.56097\; -0.07492*Age\; +\; 0.08356*1\; +\; 1.64392*1)} {(1+\; exp(1.56097\; -0.07492*Age\; +\; 0.08356*1\; +\; 1.64392*1))}
\end{equation}
\noindent The predicted probability plot for equation 18 , 19, 20 and 21 are as below;

\noindent \includegraphics*[width=4.83in, height=4.73in, keepaspectratio=false]{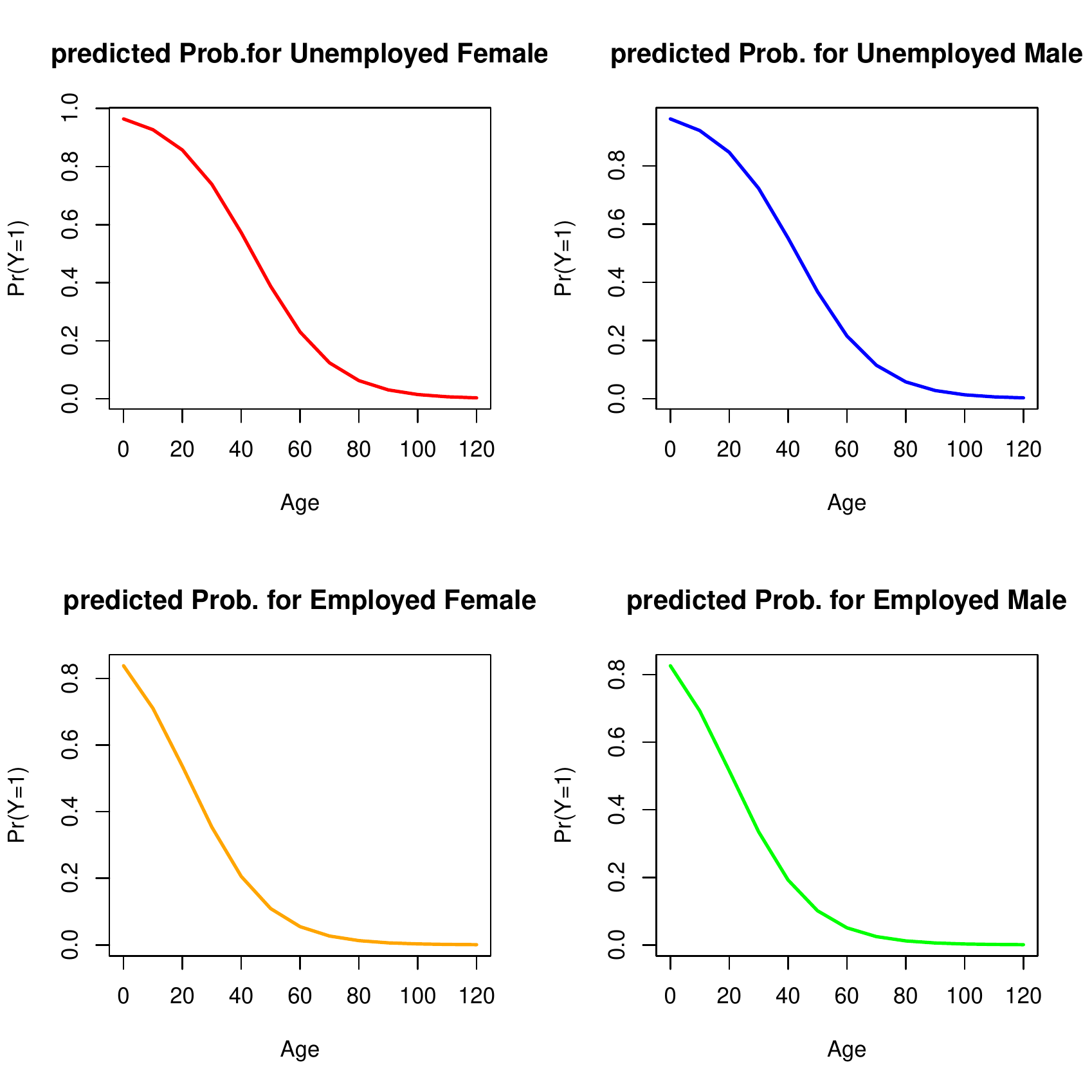}
\begin{center}
{\bf Fig 5}          
\end{center}                                                                                                          

\noindent From figure 5, the probability plot predicts all most likely of HIV contraction for female-unemployed population ageing below 39 years in Oyo state population as the fit is asymptotically approaching one. The model for the male-employed takes second position with the highest probability of HIV contraction across a randomly generated population ageing from 0 to 120 years using R sequential function (See Fig 5 for the plots).

\noindent \includegraphics*[width=6in, height=2.9in, keepaspectratio=false]{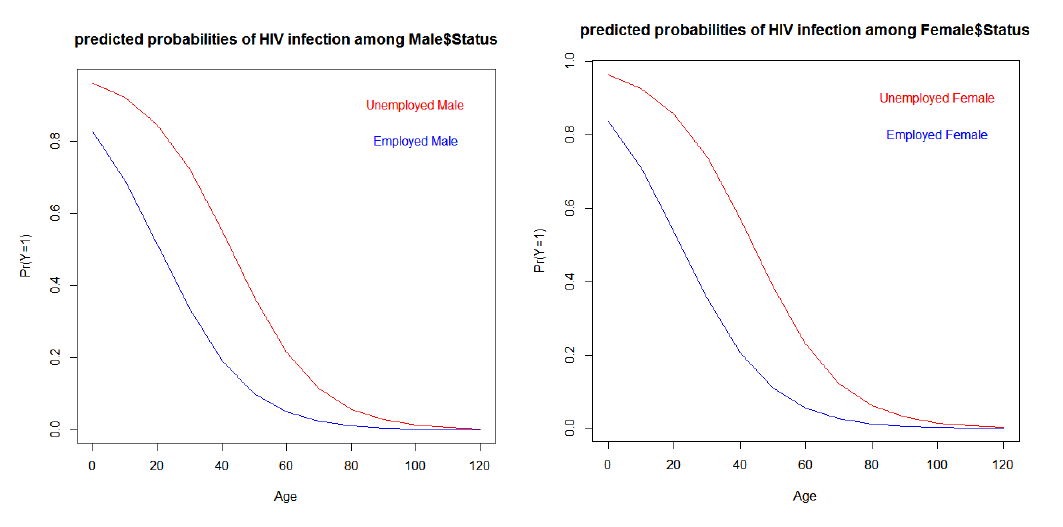}
\begin{center}
{\bf Fig 6   \& 7}          
\end{center}                                                  

\noindent For the predicted probabilities of HIV infection for females that are employed compared with females unemployed, the plot shows a increasing trend of infection with the unemployed population both in male and female gender in Oyo state, it is notably that the plot predicted the highest trend of most likelihood of infection with the unemployed females (see figure 6 and 7). 

\section{Conclusion}
\noindent The age group as block effect shows adequate significant level to HIV infection with F-value 6.496 and significant level 0.004(p$<$0.05). The age group 16-39 seems to be the age block that is most infected in the population, this age group is the reproductive age and the most sexually active stage of any population which suggests that any additional to the uncontrolled activities of sexual intercourse and pregnancy without proper medical supports will increase the cases of mother-to- chid infection in particular. An individual will not contract HIV because he/she belongs to a particular gender; contraction is majorly as result of activities or exposure. We recommend increment in employment allocation, especially for the female gender in Oyo state as a vital control measure to mitigate the spread of HIV/AIDS coupled with increase in public awareness, abstinence, and a more comprehensive approach to preventing mother-to-child infection

\noindent 

\subsection*{Reference}
\footnote{ Manuscript* of the corresponding author: oyedele.bello@futminna.edu.ng (A. O. Bello) Published online at http://journal.sapub.org/statistics Copyright © 2015 Scientific \& Academic Publishing. All Rights Reserved}
\begin{itemize}

\item [[ 1]] Aaron (2005), Logistic Regression  $http://pages.uoregon.edu/aarong/teaching/G4075_outlines/node16.html, 2005-12-21.$
\item [[ 2]] AIDS Centre\@2002-2010;\ \ http://aids.md/aids/index.php?cmd=item\&id=280\&lang\=ru
\item [[ 3]] Bello, Oyedele Adeshina; Bamiduro, Timothy Adebayo; Chuwkwu, Unna Angela $>$ Osowole, Oyedeji Isola (2015) “Bootstrap Nonlinear Regression Application in a Design of an Experiment Data for Fewer Sample Size” international journal of research (ijr) E-ISSN: 2348-6848, p- ISSN: 2348-795x volume 2, issue 2, feb. 2015 http://internationaljournalofresearch.org
\item [[ 4]] ILO$(www.ilo.org/wcmsp5/groups/public/---.../wcms_343153$.pdf.
\item [[ 5]] J. Martin Bland; Altman, Douglas G. British Medical Journal, International edition320.7247 (May 27, 2000): 1468.
\item [[ 6]] Jinma Ren, Zhen Ning, Carmen S. Kirkness, Carl V. Asche and Huaping Wang(2014) “Risk of Using logistic Regression to illustrate exposure-response relationship of infectious diseases”. BMC infectious diseases BioMed Centre Springer Science; http//www.biomedcentral.com/1471-2334/14/504\#
\item [[ 7]] L.M. Roposo, M.B. Arruda, R.M Brindeiro, F.F. Nobre; Logistic Regression Model for Predicting Resistance to HIV Protease Inhibitor Nelfinerir, XIII Mediterranean Conference on medical and Biological Engineering and Computing 2013,IFMBE proceedings volume 41, 2014,pp 1237-1240. Springer,http://link.spriger.com/chapter/10.1007\%2$F978-3-319-2_306$\#
\item [[ 8]] Scott A. Czepiel, Maximum Likelihood Estimation of Logistic Regression Models: Theory and Implementation, http://czep.net/contact.html and.
\item [[ 9]] Michael H. Kutner, Christopher J. Nachtsheim, William li; Applied Linear Statistical Models Fifth Edition, EmORY University of Minnesota John Neter University of Georgia University of Minnesota, $pp 555-581.$
\item [[10]] Open Dictionary;synonyms/Macmillandictionary.com/dictionary/British/employment Publisher Limited (2005), $www.macmillandictionary.com.$
\item [[11]] Pedhazur, E. (1997). Multiple Regression in behavioural research. New York, Harcourt Brace College Publishers: $714-765.$
\item [[12]] Richard Williams, Maximum Likelihood Estimation University of Notre Dame, $http://www3.nd.edu/~rwilliam/$ Last revised February 21, 2015.
\item [[13]] UNAIDS, 2010. UNAIDS Report on the global AIDS epidemic 2010. 

$www.unaids.org/documents/20101123_GlobalReport_em.pdf.$

\end{itemize}
\section*{Appendix}
\subsection*{The R-Programming Codes}
\begin{verbatim}
file="C:/Users/AOBELLO/Desktop/XXX/HIVtwo.csv"##GENDER, AGE(23,49,31 ETC),HIV
read.csv(file) -> HivAids.dat2
str(HivAids.dat2)
attach(HivAids.dat2)
fix(HivAids.dat2)
data.frame(HivAids.dat2)
###################################
file="C:/Users/AOBELLO/Desktop/XXX/HIVtwo50samples.csv"##GENDER, AGE(23,49,31 ETC),HIV
read.csv(file) -> HivAids.dat2
str(HivAids.dat2)
attach(HivAids.dat2)
fix(HivAids.dat2)
data.frame(HivAids.dat2)
#######################################################

file="C:/Users/AOBELLO/Desktop/datawa/HIVtwo100samples.csv"##GENDER, AGE(23,49,31 ETC),HIV
read.csv(file) -> HivAids.dat2
str(HivAids.dat2)
attach(HivAids.dat2)
fix(HivAids.dat2)
data.frame(HivAids.dat2)
############################################################

file="C:/Users/AOBELLO/Desktop/XXXX/HIVtwo300.csv"##GENDER, AGE(23,49,31 ETC),HIV
read.csv(file) -> HivAids.dat2
str(HivAids.dat2)
attach(HivAids.dat2)
fix(HivAids.dat2)
data.frame(HivAids.dat2)
#############################################################

dat=as.data.frame(cbind(Age,HIV)) # 
quartz(title="Age vs. HIV") #

plot(Age,HIV,xlab="Age",ylab="Probability of HivAids Infection" ) #
g=glm(HIV~Age,family=binomial,dat) #

curve(predict(g,data.frame(Age=x),type="resp"),add=TRUE) # draws a curve based on prediction from logistic regression model

text(70, 0.8, "50 samples")
text(70, 0.8, "100 samples")
text(70, 0.8, "300 samples")
text(70, 0.8, "400 samples")



#############################
library(rsm)
HivAids2.lm <- lm(HIV ~ poly(Sex, Age,Emp, degree=1), data=HivAids.dat2)
   #######################
par(mfrow=c(1,1))

HivAids2.lm <- lm(HIV ~ poly(Emp, Sex, degree=1), data=HivAids.dat2)
persp(HivAids2.lm, Emp ~ Sex,col = "green", zlab = "HIV")

HivAids2.lm <- lm(HIV ~ poly(Age, Emp, degree=1), data=HivAids.dat2)
persp(HivAids2.lm, Age ~ Emp,col = "green", zlab = "HIV",contours = list(z="top", col="orange"),
theta = 4, phi = 37, shade = 1)

HivAids2.lm <- lm(HIV ~ poly( Sex,Age, degree=1), data=HivAids.dat2)
persp(HivAids2.lm, Sex ~ Age,col = "green", zlab = "HIV",contours = list(z="top", col="orange"),
theta = 4, phi = 37, shade = 1)

boot.h
function(data, indices) {
 data <- data[indices, ]
 mod <- glm(formula = Kyphosis ~ Age +
 Start + Number, family = binomial, data
 = data)
 coefficients (mod)
} 
########################################################
HivAids2 <- glm(HIV ~ Age*Emp, data=HivAids.dat2, family=binomial)
HivAids2 <- glm(HIV ~ Age+Emp+Gender, data=HivAids.dat2, family=binomial(link="logit"))
summary(HivAids2)
anova(HivAids2, test="Chisq")
plot(HivAids2,which=1)
plot(HivAids2,which=2)
plot(HivAids2,which=3)
plot(HivAids2,which=4)
plot(HivAids2,which=5)
plot(HivAids2,which=6)

predict(HivAids2, type="response")

plot(predict(HivAids2),residuals(HivAids2))
abline(h=0,lty=2,col="grey")
res <- HIV(HivAids2, type = "deviance")
 
plot(predict(HivAids2), res,
xlab="Fitted values", ylab = "HIV",
ylim = max(abs(res)) * c(-1,1))
 abline(h = 0, lty = 2)

my.mod <- glm(HIV ~ Age+Emp+Gender, data=HivAids.dat2,  family = "binomial")
summary(my.mod)

# Set up the non-parametric bootstrap
logit.bootstrapHIV <- function(data, indices) {
  d <- data[indices, ]
  fit <- glm(HIV ~ Age+Emp+Gender, data = d, family = "binomial")
  return(coef(fit))
}


logit.boot <- boot(data=HivAids.dat2, statistic=logit.bootstrapHIV, R=1000) # 10'000 samples

logit.boot

# Calculate confidence intervals (Bias corrected ="bca") for each coefficient

boot.ci(logit.boot, type="bca", index=1) # intercept

boot.ci(logit.boot, type="bca", index=2) # Age

boot.ci(logit.boot, type="bca", index=3) # Emp

boot.ci(logit.boot, type="bca", index=4) # Gender
###########################################################
dat=as.data.frame(cbind(Age,HIV))
plot(Age,HIV,xlab="Age",ylab="Probability of HIV Infection") 
g=glm(HIV~Age,family=binomial,dat) # run a logistic regression model 
summary(g)
anova(g, test="Chisq")

curve(predict(g,data.frame(Age=x),type="resp"),add=TRUE) # draws a curve based on prediction from logistic regression model
points(Age,fitted(g),pch=20) 
###########################################################
dat=as.data.frame(cbind(Gender,HIV,Emp,Age))
g1=glm(HIV~Gender+Age+Emp,family=binomial,dat) # 
summary(g1)
anova(g1, test="Chisq")
res <- residuals(g1, type = "deviance")
 plot(predict(g1), res,
xlab="Fitted values", ylab = "Residuals",
ylim = max(abs(res)) * c(-1,1))
 abline(h = 0, lty = 2)

library(popbio)
logi.hist.plot(Age,HIV,boxp=FALSE,type="hist",col="gray")

#######################################################
dat=as.data.frame(cbind(Gender,HIV,Emp,Age))
g2=glm(HIV~Gender+Age*Emp,family=binomial,dat)
summary(g2)
anova(g2, test="Chisq")
confint(g2)
exp(coef(g2))
exp(confint(g2))

res <- residuals(g2, type = "deviance")
 plot(predict(g2), res,
xlab="Fitted values", ylab = "Residuals",
ylim = max(abs(res)) * c(-1,1))
 abline(h = 0, lty = 2)
##########################################
plot(g2$fitted)
abline(v=30.5,col="red")
abline(h=.5,col="blue")
abline(h=.6,col="green")
text(15,.9,"seen = 0")
text(40,.9,"seen = 1")

ggplot(HivAids.dat2, aes(x = Gender, y = HIV, colour = Age, group = Age)) + 
  geom_line()


plot(predict(g2),residuals(g2))
abline(h=0,lty=2,col="grey")
###############################################

library(popbio)
logi.hist.plot(Age,HIV,boxp=FALSE,type="hist",col="gray")
glm.out = glm(HIV ~ Gender + Emp + Age, family=binomial(logit), data=HivAids.dat2)
summary(glm.out)
Age = seq(from=0, to=120, by=10)

logOdds.M.Emp = 1.56097 -0.07492*Age
logOdds.F.Emp = 1.56097 -0.07492*Age + 0.08356*1
logOdds.M.UnEmp   = 1.56097 -0.07492*Age             + 1.64392*1
logOdds.F.UnEmp   = 1.56097 -0.07492*Age + 0.08356*1 + 1.64392*1

prob.logOdds.M.Emp = exp(logOdds.M.Emp)/(1+ exp(logOdds.M.Emp))
prob.logOdds.F.Emp = exp(logOdds.F.Emp)/(1+ exp(logOdds.F.Emp))
prob.logOdds.M.UnEmp   = exp(logOdds.M.UnEmp)  /(1+ exp(logOdds.M.UnEmp))
prob.logOdds.F.UnEmp   = exp(logOdds.F.UnEmp)  /(1+ exp(logOdds.F.UnEmp))

windows()
  par(mfrow=c(2,2))
  plot(x=Age, y=prob.logOdds.F.UnEmp, type="l", col="red", lwd=2, 
       ylab="Pr(Y=1)", main="predicted probabilities for Unemployed Female")
  plot(x=Age, y=prob.logOdds.M.UnEmp , type="l", col="blue", lwd=2, 
       ylab="Pr(Y=1)", main="predicted probabilities for Unemployed Male")
  plot(x=Age, y=prob.logOdds.F.Emp, type="l", col="orange", lwd=2, 
       ylab="Pr(Y=1)", main="predicted probabilities for Employed Female")
  plot(x=Age, y=prob.logOdds.M.Emp, type="l", col="green", lwd=2, 
       ylab="Pr(Y=1)", main="predicted probabilities for Employed Male")

windows()
  plot(x=Age, y=prob.logOdds.F.UnEmp, type="l", col="red", lwd=1, 
       ylab="Pr(Y=1)", main="predicted probabilities of HIV infection among Female$Status")
  lines(x=Age, y=prob.logOdds.F.Emp, col="blue", lwd=1)
text(100, 0.9, "Unemployed Female", col="red")
text(100, 0.8, "Employed Female",col="blue")

windows()
  plot(x=Age, y=prob.logOdds.M.UnEmp, type="l", col="red", lwd=1, 
       ylab="Pr(Y=1)", main="predicted probabilities of HIV infection among Male$Status")
  lines(x=Age, y=prob.logOdds.M.Emp, col="blue", lwd=1)
text(100, 0.9, "Unemployed Male", col="red")
text(100, 0.8, "Employed Male",col="blue")

legend("topright", legend=col


windows()
  plot(x=Age, y=prob.logOdds.F.UnEmp, type="l", col="red", lwd=1, 
       ylab="Pr(Y=1)", main="predicted probabilities of HIV infection")
  lines(x=Age, y=prob.logOdds.M.UnEmp, col="blue", lwd=1)
d;pd;;de;
         "Employed Female", "Employed Male"), lty=c("solid", "solid", "dotted",
         "dotted"), lwd=c(1,1,2,2), col=c("red", "blue", "red", "blue"))
##################################################################################
Validity#######
file="C:/Users/AOBELLO/Desktop/XXXX/validate.csv"##GENDER, AGE(23,49,31 ETC),HIV
read.csv(file) -> HivAids.dat2
str(HivAids.dat2)
attach(HivAids.dat2)
fix(HivAids.dat2)
data.frame(HivAids.dat2)

HivAids.model   = 1.30131 -0.06921*Age + 0.15188*Gender + 1.67899*Emp

HIV
prob.Odds.HivAids.model = exp(HivAids.model)/(1+ exp(HivAids.model))
round(prob.Odds.HivAids.model)
round(prob.Odds.HivAids.model, digit=0)
signif(prob.Odds.HivAids.model,digit=0)
\end{verbatim}
\end{document}